\documentstyle[prd,aps,floats,psfig]{revtex}
\begin{document}
\draft
\title{``No--Hair'' Theorem for Spontaneously Broken Abelian Models in Static Black
Holes}
\author{Eloy Ay\'{o}n--Beato}
\address{Departamento~de~F\'{\i}sica,~CINVESTAV--IPN,\\
Apartado~Postal~14--740,~C.P.~07000,~M\'{e}xico,~D.F.,~MEXICO.}
\maketitle

\begin{abstract}
The vanishing of the electromagnetic field, for purely electric
configurations of spontaneously broken Abelian models, is established in the
domain of outer communications of a static asymptotically flat black hole.
The proof is gauge invariant, and is accomplished without any dependence on
the model. In the particular case of the Abelian Higgs model, it is shown
that the only solutions admitted for the scalar field become the vacuum
expectation values of the self--interaction.
\end{abstract}


\preprint{gr-qc/9611069}

\section{Introduction}

\label{sec:intro}

The classical and strongest version of the ``no--hair'' conjecture
establishes that a stationary black hole is uniquely described by global
charges, i.e., conserved charges associated with massless gauge fields,
expressed by surface integrals at the spatial infinity $i^{o}$ \cite{Bizon94}%
. In particular, the conjecture excludes the existence of massive fields in
the domain of outer communications $\langle \!\langle
\hbox{${\cal J}$\kern -.645em
{\raise.57ex\hbox{$\scriptscriptstyle (\ $}}}\rangle \!\rangle $ of a
stationary black hole. This fact rests on the idea that in the black hole
transition to stationarity ``everything that can be radiated away will be
radiated away'' (cf.~\cite{Price72}), so, the only classical degrees of
freedom of a stationary black hole are those corresponding to non--radiative
multipole moments; massive fields are automatically excluded because all
their multipoles are radiative \cite{Bizon94}.

The absence of massive ``hair'' was shown early in the Bekenstein pioneering
works for massive scalar fields, Proca--massive spin--1 fields, and massive
spin--2 fields \cite{Bekens72a,Bekens72b,Bekens72c}. An alternative
demonstration for Proca fields can be found in \cite{Ayon99a}. The
``no--hair'' theorem for massive vector fields is a useful tool for
excluding the existence of new black hole solutions for very complicated
theories as metric--affine gravity, where a relevant sector of this theory
reduces to an effective Einstein--Proca system \cite{AyonGMQ99}.

It is well--known that fields acquire mass not only kinematically, as in the
previous cases, but also through a dynamical mechanism of spontaneous
symmetry breaking. This is the case of spontaneously broken Abelian models
describing a charged scalar field with a self--interaction having nonzero
vacuum expectation values, and minimally coupled to a massless Abelian gauge
field. The ``no--hair'' conjecture for these models has been previously
enunciated as \cite{C-P-W}: {\em any stationary black hole solution, such
that all gauge--invariant observables are non--singular, must have a
vanishing electromagnetic field, in the domain of outer communications $%
\langle \!\langle
\hbox{${\cal J}$\kern -.645em
{\raise.57ex\hbox{$\scriptscriptstyle (\ $}}}\rangle \!\rangle $ of the
black hole}. The simplest of this systems is the Abelian Higgs model
(Mexican--hat self--interaction) for which a ``no--hair'' theorem was shown
in \cite{AdlerP79}, proving the vanishing of the gauge field for spherically
symmetric static black holes. This proof has been considered unsatisfactory
\cite{Gibbons91} because it is based on an inconsistent gauge choice.
Improved versions have been recently given \cite
{Lahiri92,MayoBeke96,Bekens96}, without the original restrictions criticized
in \cite{Gibbons91}.

The subject of this paper is twofold, first, to relax the spherically
symmetric assumption in the previously quoted contributions, by working with
general static asymptotically flat systems, and second, to extent the
``no--hair'' theorem to more general Abelian models than the Higgs model,
i.e., for general spontaneously broken self--interactions. Emphasis is given
on asymptotically flat black holes only, this way we exclude from
consideration black holes pierced by a cosmic string \cite{AchucarGK95}
---with the corresponding nontrivial behavior of the Abelian field---, as it
has been previously pointed by Bekenstein \cite{Bekens96,Bekens98}, these
last configurations are not asymptotically flat since they present the
angular deficit inherent to the presence of topological defects. The basic
difference between these configurations is that for the string--pierced
black holes the scalar field satisfy boundary conditions at infinity in
accordance with the existence of a topological defect, i.e., the scalar
field is confined to the vacuum only in a circle at infinity, which implies
the developing of a cosmic string at the interior of the circle, whereas for
asymptotically flat black holes the scalar field approaches the vacuum in
all directions at infinity.%

For a static black hole, the Killing field {\boldmath$k$} coincides with the
null generator of the event horizon ${{\cal {H}}}^{+}$ and is timelike and
hypersurface orthogonal in all the domain of outer communications $\langle
\!\langle
\hbox{${\cal J}$\kern -.645em
{\raise.57ex\hbox{$\scriptscriptstyle (\ $}}}\rangle \!\rangle $. This allow
us to choose, by simply connectedness of $\langle \!\langle
\hbox{${\cal J}$\kern -.645em
{\raise.57ex\hbox{$\scriptscriptstyle (\ $}}}\rangle \!\rangle $ \cite
{ChrWald95}, a global coordinate system $(t,x^i)$, $i=1,2,3$, in all $%
\langle \!\langle
\hbox{${\cal J}$\kern -.645em
{\raise.57ex\hbox{$\scriptscriptstyle (\ $}}}\rangle \!\rangle $ \cite
{Carter87}, such that $\text{\boldmath$k$}=\text{\boldmath$\partial
/\partial t$}$ and the metric reads
\begin{equation}
\text{\boldmath$g$}=-V\text{\boldmath$dt$}^2+\gamma _{ij}\text{\boldmath$dx$}%
^i\text{\boldmath$dx$}^j,  \label{eq:static}
\end{equation}
where $V$ and {\boldmath$\gamma $} are $t$--independent, {\boldmath$\gamma $}
is positive definite in all $\langle \!\langle
\hbox{${\cal J}$\kern -.645em
{\raise.57ex\hbox{$\scriptscriptstyle(\ $}}}\rangle \!\rangle $, and $V$ is
positive in all $\langle \!\langle
\hbox{${\cal J}$\kern -.645em
{\raise.57ex\hbox{$\scriptscriptstyle(\ $}}}\rangle \!\rangle $ and vanishes
in ${{\cal {H}}}^{+}$. From (\ref{eq:static}) it can be noticed that
staticity implies the existence of a time--reversal isometry $t\mapsto -t$.

In the next Sec.~\ref{sec:gauge} the vanishing of the electromagnetic field
in the domain of outer communications $\langle \!\langle
\hbox{${\cal J}$\kern -.645em
{\raise.57ex\hbox{$\scriptscriptstyle (\ $}}}\rangle \!\rangle $ of a static
asymptotically flat black hole is demonstrated for purely electric
configurations of a generic spontaneously broken Abelian model. At the end
of Sec.~\ref{sec:gauge} the conditions for establishing a ``no--hair''
theorem for purely magnetic configurations are also analyzed. Sec.~\ref
{sec:scalar} is devoted to show, in the particular case of the Abelian Higgs
model, that the charged scalar field is confined to its vacuum in $\langle
\!\langle
\hbox{${\cal J}$\kern -.645em
{\raise.57ex\hbox{$\scriptscriptstyle (\ $}}}\rangle \!\rangle $.
Conclusions are given in Sec.~\ref{sec:conclu}.

\section{``No--Hair'' Theorem for the Abelian Gauge Field}

\label{sec:gauge}

\begin{figure}[h]
\centerline{\psfig{file=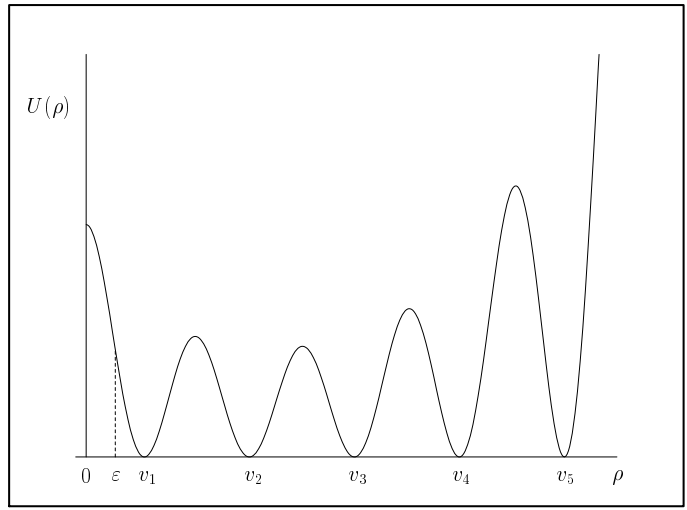,width=10cm}}
\caption{Example of a spontaneously broken potential with five types of
non--vanishing vacuum expectation values. The positive real number $%
\varepsilon $ is such that $0<\varepsilon \leq v_1$, and it will be used to
show that $\rho $ is a non--vanishing function at the horizon.}
\label{fig:poten}
\end{figure}

The action describing the coupling to gravity of the relevant models to be
considered is
\begin{equation}
{\cal S}=\frac 12\int \left( \frac 1\kappa R-\frac 1{8\pi }F_{\alpha \beta
}F^{\alpha \beta }-(D_\alpha \Phi )^{\dagger }D^\alpha \Phi -U(\Phi
^{\dagger }\Phi )\right) dv,  \label{eq:L}
\end{equation}
where $R$ is the scalar curvature, $F_{\alpha \beta }\equiv 2\nabla
_{[\alpha }A_{\beta ]}$ is the field strength of the Abelian gauge field $%
A_\alpha $, $D_\alpha \equiv \nabla _\alpha -ieA_\alpha $ is the gauge
covariant derivative, and $U(\Phi ^{\dagger }\Phi )$ is a non--negative
self--interaction with nonvanishing vacuum expectation values, as for
instance, in the Higgs model where $U_{\text{H}}=(\lambda /2)(|\Phi
|^2-v^2)^2$; here $(\cdot )^{\dagger }$ denotes complex conjugation.
Parametrizing the complex scalar field by $\Phi =\rho \exp i\theta $ the
Lagrangian becomes
\begin{equation}
L=\frac 1{2\kappa }R-\frac 1{16\pi }F_{\alpha \beta }F^{\alpha \beta }-\frac
12\nabla _\alpha \rho \nabla ^\alpha \rho -\frac 1{2e^2\rho ^2}J_\alpha
J^\alpha -\frac 12U(\rho ),  \label{eq:Lp}
\end{equation}
with $J_\alpha \equiv e\rho ^2(\nabla _\alpha \theta -eA_\alpha )$. The
potential $U(\rho )$ is a non--negative function achieving its minima at
nonzero values $v_a$, cf.~Fig.~\ref{fig:poten}, and it is assumed that $\rho
$ asymptotically approaches to any one of this values.
The Abelian symmetry of the models is expressed by the invariance of the
Lagrangian (\ref{eq:Lp}) under the gauge transformations $\theta \mapsto
\theta +e\Lambda $, $A_\alpha \mapsto A_\alpha +\nabla _\alpha \Lambda $.
From the Lagrangian (\ref{eq:Lp}), the Einstein--Maxwell--Scalar equations
for the involved fields are established
\begin{equation}
\frac 1\kappa R_{\mu \nu }=\frac 1{4\pi }F_\mu ^{~\alpha }F_{\nu \alpha
}+\nabla _\mu \rho \nabla _\nu \rho +\frac 1{e^2\rho ^2}J_\mu J_\nu -\frac 12%
g_{\mu \nu }\left( \frac{F_{\alpha \beta }F^{\alpha \beta }}{8\pi }-U(\rho
)\right) ,  \label{eq:Ein}
\end{equation}
\begin{equation}
\nabla _\beta F^{\alpha \beta }=4\pi J^\alpha ,  \label{eq:Max}
\end{equation}
\begin{equation}
\Box \rho =\frac 12U^{\prime }(\rho )+\frac 1{e^2\rho ^3}J_\alpha J^\alpha ,
\label{eq:scalar}
\end{equation}
where $U^{\prime }(\rho )\equiv dU(\rho )/d\rho $.

We would like to emphasize that the Reissner--Nordstr\"om black hole is not
a solution of the above equations; the system we are dealing with is an
Abelian Higgs model, i.e., a charged ($e\neq 0$) scalar field minimally
coupled to an Abelian gauge field, and with a self--interaction having
nonvanishing vacuum expectation values. The coupling of this system to
gravity (\ref{eq:Ein})--(\ref{eq:scalar}), does not reduce in no one case to
the Einstein--Maxwell system, and therefore, it does not contain the
Reissner--Nordstr\"om black hole as a solution. This becomes apparent from
the Lagrangian (\ref{eq:L}): for constant values of the charged scalar
field, $\Phi =\text{{\rm const.}}$, a mass term, $e^2|\text{{\rm const.}}%
|^2A_\mu A^\mu $, is present, which converts the Abelian gauge field in a
massive Proca--like spin--1 field, for which there exist no static black
hole solutions except the Schwarzschild one, as it was pointed out in the
introduction \cite{Bekens72b,Ayon99a}. For a zero value of the scalar field,
the mass term vanishes, but, an effective cosmological constant, $\Lambda _{%
{\rm eff}}=\kappa U(0)/2$, arises, this is due to the spontaneously broken
behavior of the self--interaction, which requires $U(0)\neq 0$. In this
case, we lost asymptotic flatness and, consequently, the
Reissner--Nordstr\"om black hole cannot be a solution of the resulting
system. Other is the situation when there is no spontaneously symmetry
breaking, i.e., $U(0)=0$, in this case the model reduces to the
Einstein--Maxwell system for vanishing scalar field and the existence of the
Reissner--Nordstr\"om black hole is assured, but this is not the case we
will dealt with in the paper.

We shall assume that the gauge field shares the same symmetries of the
metric, namely, it is stationary, $\text{\boldmath${\pounds }_kF$}=0$.
Consequently with a (metric--)static configuration (\ref{eq:static}), we
will also assume the existence of electromagnetic staticity, i.e., the
Maxwell field $F^{\alpha \beta }$ and the Maxwell equations (\ref{eq:Max})
are invariant under time--reversal transformations. The time--reversal
invariance of Maxwell equations (\ref{eq:Max}) requires that, in the
coordinates chosen in (\ref{eq:static}), $J^t$ and $F^{ti}$ remain unchanged
while $J^i$ and $F^{ij}$ change sign, or the opposite scheme, i.e., $J^t$
and $F^{ti}$ change sign as long as $J^i$ and $F^{ij}$ remain unchanged
under time reversal \cite{Bekens72b}. However, this isometry should not
change gauge--invariant observables, therefore $J^i$ and $F^{ij}$ must
vanish in the first case, while $J^t$ and $F^{ti}$ vanish in the second one.
Hence, staticity on the metric and material sources implies the existence of
two nonoverlapping cases: a purely electric case (I) and a purely magnetic
case (II).

Now we are ready to proof the ``no--hair'' statement for the gauge field,
i.e., for spontaneously broken Abelian model the electromagnetic field
vanishes in the domain of outer communications $\langle \!\langle
\hbox{${\cal J}$\kern -.645em
{\raise.57ex\hbox{$\scriptscriptstyle (\ $}}}\rangle \!\rangle $ of a static
asymptotically flat black hole. Let ${\cal V}\subset \langle \!\langle
\hbox{${\cal J}$\kern -.645em
{\raise.57ex\hbox{$\scriptscriptstyle (\ $}}}\rangle \!\rangle $ be the open
region bounded by the spacelike hypersurface $\Sigma $, the spacelike
hypersurface $\Sigma ^{\prime }$, and pertinent portions of the horizon $%
{\cal H}^{+}$, and the spatial infinity $i^o$. The spacelike hypersurface $%
\Sigma ^{\prime }$ is obtained by shifting each point of $\Sigma $ a unit
parametric value along the integral curves of the Killing field {\boldmath$k$%
}. Multiplying the Maxwell equations (\ref{eq:Max}) by $J_\alpha /\rho ^2$
and integrating by parts over ${\cal {V}}$, after applying the Gauss
theorem, and using that $J_\alpha /\rho ^2=2e(\nabla _\alpha \theta
-eA_\alpha )$, one obtains
\begin{equation}
\left[ \int_{\Sigma ^{\prime }}-\int_\Sigma +\int_{{\cal {H}}^{+}\cap
\overline{{\cal {V}}}}+\int_{i^o\cap \overline{{\cal {V}}}}\right] \frac 1{%
\rho ^2}J_\alpha F^{\alpha \beta }d\Sigma _\beta =\int_{{\cal {V}}}\left(
\frac{e^2}2F_{\alpha \beta }F^{\alpha \beta }+\frac{4\pi }{\rho ^2}J_\alpha
J^\alpha \right) dv,  \label{eq:int}
\end{equation}
The boundary integral over $\Sigma ^{\prime }$ cancels out that one over $%
\Sigma $, since $\Sigma ^{\prime }$ and $\Sigma $ are isometric
hypersurfaces. At spatial infinity $i^o$ the scalar field modulus $\rho $
approaches to one of the nonvanishing values, $v_a$, minimizing the
potential function $U(\rho )$, which implies (see the Lagrangian (\ref{eq:Lp}%
)) that the gauge field behaves as an effective massive field at spatial
infinity $i^o$, due to the spontaneous breaking of the gauge symmetry at
this region. The usual Yukawa fall--off of massive fields at infinity cause
the boundary integral over $i^o\cap \overline{{\cal {V}}}$ vanishes \cite
{AdlerP79,Lahiri92}. For the remaining boundary integral at the portion of
the horizon ${\cal {H}}^{+}\cap \overline{{\cal {V}}}$ we make use of the
standard measure on this region $d\,\Sigma _\beta =2n_{[\beta }l_{\mu
]}l^\mu d\sigma $ \cite{Zannias95,Zannias98}, where {\boldmath$l$} is the
null generator of the horizon, {\boldmath$n$} is the other future--directed
null vector ($n_\mu l^\mu =-1$), orthogonal to the spacelike cross sections
of the horizon, and $d\sigma $ is the surface element ---the standard
measure follows from choosing the natural volume 3--form at the horizon,
i.e., $\text{\boldmath$\eta _3$}={}^{*}(\text{\boldmath$n$}\wedge \text{%
\boldmath$l$})\wedge \text{\boldmath$l$}$. Using the quoted measure the
integrand over the horizon can be rewritten as
\begin{equation}
\frac 1{\rho ^2}J_\alpha F^{\alpha \beta }d\Sigma _\beta =\left( \frac{%
J_\alpha F^{\alpha \beta }l_\beta }{\rho ^2}+\frac{J_\alpha F^{\alpha \beta
}n_\beta }{\rho ^2}\left( l_\mu l^\mu \right) \right) d\sigma \,.
\label{eq:integ}
\end{equation}
In order to demonstrate that the last integrand vanishes it is sufficient to
prove that the quantities appearing at the right hand side of (\ref{eq:integ}%
) are such that: $J_\alpha F^{\alpha \beta }l_\beta /\rho ^2$ vanishes and $%
J_\alpha F^{\alpha \beta }n_\beta /\rho ^2$ remains bounded at the horizon.
We shall establish the behavior of these quantities at the horizon by
studying some invariants constructed from the curvature. By using Einstein
equations (\ref{eq:Ein}), we obtain the following two equivalent
expressions,
\begin{eqnarray}
\frac 2{\kappa ^2}R_{\mu \nu }R^{\mu \nu } &=&\frac{G^2}{2\pi }+\left( \frac{%
2J^\mu \nabla _\mu \rho }{e\rho }\right) ^2+\left( \frac F{2\pi }-\frac{%
J_\mu J^\mu }{e^2\rho ^2}\right) ^2+\left( \frac F{2\pi }+\nabla _\mu \rho
\nabla ^\mu \rho \right) ^2+\left( U(\rho )+\frac{J_\mu J^\mu }{e^2\rho ^2}%
\right) ^2  \nonumber \\
&&\ \ +(U(\rho )+\nabla _\mu \rho \nabla ^\mu \rho )^2+\frac 1{\pi e^2\rho ^2%
}J^\mu F_\mu ^{~\alpha }J^\nu F_{\nu \alpha }+\frac 1\pi \nabla ^\mu \rho
\,{}^{*}\!F_\mu ^{~\alpha }\nabla ^\nu \rho \,{}^{*}\!F_{\nu \alpha },
\label{eq:RicciSE} \\
\ &=&\frac{G^2}{2\pi }+\left( \frac{2J^\mu \nabla _\mu \rho }{e\rho }\right)
^2+\left( \frac F{2\pi }-\frac{J_\mu J^\mu }{e^2\rho ^2}\right) ^2+\left(
\frac F{2\pi }-\nabla _\mu \rho \nabla ^\mu \rho \right) ^2+\left( U(\rho )+%
\frac{J_\mu J^\mu }{e^2\rho ^2}\right) ^2  \nonumber \\
&&\ \ +(U(\rho )+\nabla _\mu \rho \nabla ^\mu \rho )^2+\frac 1{\pi e^2\rho ^2%
}J^\mu F_\mu ^{~\alpha }J^\nu F_{\nu \alpha }+\frac 1\pi \nabla ^\mu \rho
F_\mu ^{~\alpha }\nabla ^\nu \rho F_{\nu \alpha },  \label{eq:RicciSM}
\end{eqnarray}
where $F\equiv F_{\alpha \beta }F^{\alpha \beta }/4$, $G\equiv
{}^{*}\!F_{\alpha \beta }F^{\alpha \beta }/4$, and ${}^{*}\!F_{\alpha \beta
} $ stands for the Hodge dual (${}^{*}\!F_{\alpha \beta }=\eta _{\mu \nu
\alpha \beta }F^{\mu \nu }/2$). It is important to note that the previous
Eqs.\ only differ in the sign inside the fourth term, and in the fact that
the last term is written in each case with ${}^{*}\!F_{\alpha \beta }$ and $%
F_{\alpha \beta }$, respectively.

Since the horizon is a smooth surface, the left hand side of the above Eqs.$%
~ $is bounded on it. For the purely electric case (I), the last two terms in
the right hand side of (\ref{eq:RicciSE}) are non--negative, the remaining
terms are perfect square, and consequently each term in the right hand side
of (\ref{eq:RicciSE}) is bounded at the horizon. In particular, the bounded
behavior of the sixth term involving the quantities $U(\rho )$ and $\nabla
_\mu \rho \nabla ^\mu \rho $ implies, from the non--negativeness of these
quantities, that they are also bounded at the horizon. It follows from the
bounded behavior of the perfect--square terms where $U(\rho )$ and $\nabla
_\mu \rho \nabla ^\mu \rho $ are combined with the quantities $J_\mu J^\mu
/e^2\rho ^2$ and $F$, respectively, that the last mentioned quantities are
also bounded at the horizon. Thus, any quantity appearing in the right hand
side of (\ref{eq:RicciSE}) is bounded at the horizon, in particular $U(\rho
) $, $F$ and $J_\mu J^\mu /\rho ^2$. The same conclusions can be achieved,
along the same lines of reasoning, for the purely magnetic case (II), but
this time using the right hand side of Eq.$~$(\ref{eq:RicciSM}). Other
invariants can be built from the Ricci curvature (\ref{eq:Ein}) by means of {%
\boldmath$l$} and {\boldmath$n$}, which are well--defined smooth vector
fields on the horizon. The first invariant reads
\begin{equation}
\frac 1\kappa R_{\mu \nu }n^\mu n^\nu =\frac 1{4\pi }I_\mu I^\mu +\left(
n^\mu \nabla _\mu \rho \right) ^2+\frac 1{e^2\rho ^2}\left( J_\mu n^\mu
\right) ^2-\frac{n_\mu n^\mu }2\left( \frac F{2\pi }-U(\rho )\right) ,
\label{eq:Riccinn}
\end{equation}
where $I^\mu \equiv F^{\mu \nu }n_\nu $. The last term above vanishes
because the bounded behavior of both the invariant $F$ and the potential $%
U(\rho )$. Since {\boldmath$I$} is orthogonal to the null vector {\boldmath$%
n $}, it must be spacelike or null ($I_\mu I^\mu \geq 0$), therefore each
one of the remaining terms on the right hand side of (\ref{eq:Riccinn}) must
be bounded. The next invariant to be considered, which vanishes at the
horizon due to the Raychaudhuri equation for the null generator \cite{Wald},
reads
\begin{equation}
0=\frac 1\kappa R_{\mu \nu }l^\mu l^\nu =\frac 1{4\pi }E_\mu E^\mu +\left(
l^\mu \nabla _\mu \rho \right) ^2+\frac 1{e^2\rho ^2}\left( J_\mu l^\mu
\right) ^2-\frac{l_\mu l^\mu }2\left( \frac F{2\pi }-U(\rho )\right) ,
\label{eq:Riccill}
\end{equation}
where $E^\mu \equiv F^{\mu \nu }l_\nu $ is the electric field at the
horizon. Once again the bounded behavior of the invariant $F$ and the
potential $U(\rho )$ can be used to achieve the vanishing of the last term
of (\ref{eq:Riccill}). Since {\boldmath$E$} is orthogonal to the null
generator {\boldmath$l$}, it must be spacelike or null ($E_\mu E^\mu \geq 0$%
), consequently each term on the right hand side of (\ref{eq:Riccill})
vanishes independently, which implies that $J_\mu l^\mu /\rho =0$ and that {%
\boldmath$E$} is proportional to the null generator {\boldmath$l$} at the
horizon, i.e., $\text{\boldmath$E$}=-(E_\alpha n^\alpha )\,\text{\boldmath$l$%
}$. The vanishing of $l^\mu \nabla _\mu \rho $, only reproduces the fact
that {\boldmath$l$} coincides with the Killing field at the horizon. The
last invariant to be studied gives the following relation:
\begin{equation}
\frac 1\kappa R_{\mu \nu }l^\mu n^\nu -\frac F{4\pi }+\frac{U(\rho )}2=\frac
1{4\pi }(E_\mu n^\mu )^2+\left( l^\mu \nabla _\mu \rho \right) \left( n^\nu
\nabla _\nu \rho \right) +\left( \frac{J_\mu l^\mu }{e\rho }\right) \left(
\frac{J_\nu n^\nu }{e\rho }\right) ,  \label{eq:Ricciln}
\end{equation}
where it has been used that $\text{\boldmath$E$}=-(E_\alpha n^\alpha )\,%
\text{\boldmath$l$}$. Because $n^\nu \nabla _\nu \rho $ and $J_\nu n^\nu
/\rho $ are bounded at the horizon, and $J_\mu l^\mu /\rho =0=l^\mu \nabla
_\mu \rho $, the last two terms in the right hand side of (\ref{eq:Ricciln})
vanish, thus $E_\mu n^\mu $ is bounded at the horizon as consequence of the
bounded behavior of the left hand side of (\ref{eq:Ricciln}).

Summarizing, the study of the quoted invariants at the horizon leads to the
following conclusions: $E_{\mu }n^{\mu }$, $J_{\mu }n^{\mu }/\rho $, $n^{\mu
}\nabla _{\mu }\rho $, $J_{\mu }J^{\mu }/\rho ^{2}$, and $I_{\mu }I^{\mu }$
are bounded at the horizon, while $J_{\mu }l^{\mu }/\rho =0$ and $\text{%
\boldmath$E$}=-(E_{\alpha }n^{\alpha })\,\text{\boldmath$l$}$ in the same
region.

Now we are in position to make a more detailed analysis of the sufficient
conditions for the vanishing of the integrand (\ref{eq:integ}) over the
horizon, i.e., $J_\alpha F^{\alpha \beta }l_\beta /\rho ^2$ vanishes and $%
J_\alpha F^{\alpha \beta }n_\beta /\rho ^2$ remains bounded at the horizon.
Using the definition $E^\mu \equiv F^{\mu \nu }l_\nu $ and that $\text{%
\boldmath$E$}=-(E_\alpha n^\alpha )\,\text{\boldmath$l$}$, we obtain for the
first quantity at the horizon
\begin{equation}
\frac{J_\alpha F^{\alpha \beta }l_\beta }{\rho ^2}=-\frac 1\rho \left( E_\mu
n^\mu \right) \frac{J_\nu l^\nu }\rho .  \label{eq:int1v}
\end{equation}
Since $E_\mu n^\mu $ is bounded and $J_\nu l^\nu /\rho $ vanishes at the
horizon, it follows that the last expression vanishes at the horizon if the
scalar field modulus $\rho $ does not vanishes in this region. We shall show
at the end of this section that $\rho $ is a nonvanishing function at the
horizon and in all the domain of outer communications $\langle \!\langle
\hbox{${\cal J}$\kern -.645em
{\raise.57ex\hbox{$\scriptscriptstyle (\ $}}}\rangle \!\rangle $.

For the second quantity we note that {\boldmath$J$} and {\boldmath$I$} are
orthogonal to the null vectors {\boldmath$l$} and {\boldmath$n$},
respectively. Therefore, {\boldmath$J$} must be spacelike or proportional to
{\boldmath$l$}, and {\boldmath$I$} must be spacelike or proportional to {%
\boldmath$n$}. Using a null tetrad basis, constructed with {\boldmath$l$}, {%
\boldmath$n$}, and a pair of linearly independent spacelike vectors, these
last ones being tangent to the spacelike cross sections of the horizon, the {%
\boldmath$J$} and {\boldmath$I$} vectors can be written as
\begin{equation}
\text{\boldmath$J$}=-(J_\alpha n^\alpha )\text{\boldmath$l$}+\text{\boldmath$%
J$}^{\bot },  \label{eq:Jnt}
\end{equation}
\begin{equation}
\text{\boldmath$I$}=-(I_\alpha l^\alpha )\text{\boldmath$n$}+\text{\boldmath$%
I$}^{\bot },  \label{eq:Int}
\end{equation}
where $\text{\boldmath$J$}^{\bot }$ and $\text{\boldmath$I$}^{\bot }$ are
the projections, orthogonal to {\boldmath$l$} and {\boldmath$n$}, on the
spacelike cross sections of the horizon. Using (\ref{eq:Jnt}) and (\ref
{eq:Int}) it is clear that $J_\mu J^\mu =J_\mu ^{\bot }J^{\bot \mu }$, and $%
I_\mu I^\mu =I_\mu ^{\bot }I^{\bot \mu }$, i.e., the contribution to these
bounded magnitudes comes only from the spacelike sector orthogonal to {%
\boldmath$l$} and {\boldmath$n$}. With the help of (\ref{eq:Jnt}) and (\ref
{eq:Int}) the other quantity appearing in the integrand (\ref{eq:integ}) can
be written as
\begin{equation}
\frac{J_\alpha F^{\alpha \beta }n_\beta }{\rho ^2}=\frac{J_\alpha I^\alpha }{%
\rho ^2}=\frac 1\rho \left\{ \left( E_\alpha n^\alpha \right) {}\frac{%
J_\beta n^\beta }\rho {}+\frac{J_\alpha ^{\bot }I^{\bot \alpha }}\rho
\right\} ,  \label{eq:int2b}
\end{equation}
where the identity $I_\alpha l^\alpha =-E_\alpha n^\alpha $ has been used.
The first term inside the braces in (\ref{eq:int2b}) is bounded because $%
E_\alpha n^\alpha $ and $J_\beta n^\beta /\rho $ are bounded. To the second
term the Schwarz inequality applies, since $\text{\boldmath$J$}^{\bot }$ and
$\text{\boldmath$I$}^{\bot }$ belong to a spacelike subspace. Thus, $%
(J_\alpha ^{\bot }I^{\bot \alpha }/\rho )^2\leq (J_\mu ^{\bot }J^{\bot \mu
}/\rho ^2)(I_\nu ^{\bot }I^{\bot \nu })=(J_\mu J^\mu /\rho ^2)(I_\nu I^\nu )$
and since $J_\mu J^\mu /\rho ^2$ and $I_\nu I^\nu $ are bounded at the
horizon, the second term inside the braces of (\ref{eq:int2b}) is also
bounded. Since the term enclosed by the braces in (\ref{eq:int2b}) is
bounded, it follows that the bounded behavior at the horizon of the whole
expression depends again in the nonvanishing property of the scalar field
modulus $\rho $ in this region.

The analysis of the sufficient conditions for the vanishing of the integrand
(\ref{eq:integ}) over the horizon shows that, the quantity (\ref{eq:int1v})
vanishes at the horizon and the quantity (\ref{eq:int2b}) remains bounded in
this region if the scalar field modulus $\rho $ is a nonvanishing function
at the horizon. All the conclusions achieved up to now, can be applied to
both cases the purely electric (I) and the purely magnetic (II) ones. To
finish the demonstration of the vanishing of the integrand (\ref{eq:integ})
over the horizon, it remains only to show that $\rho $ is a nonvanishing
function in this region. We are able, by using the function $f_\varepsilon $
below, to complete the demonstration for the purely electric case (I).
Unfortunately, the purely magnetic case (II) escapes to be treated along a
similar way and it remains still as an open problem; we believe that the
``no--hair'' conjecture applies also to this case.

\begin{figure}[h]
\centerline{\psfig{file=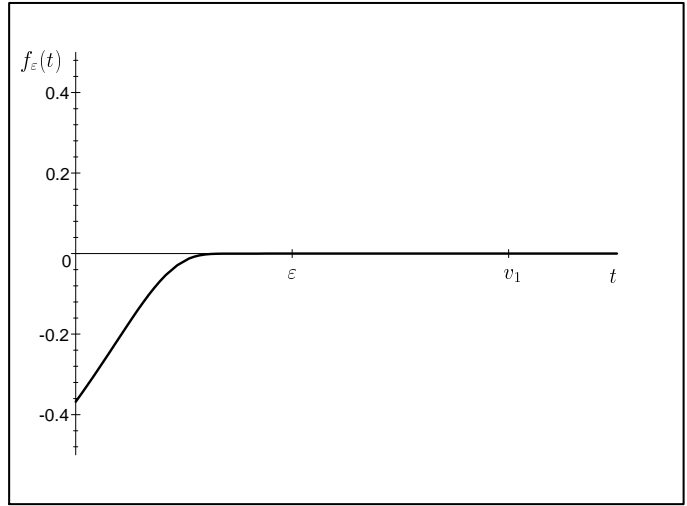,width=10cm}}
\caption{The graph of the auxiliar function $f_{\varepsilon }(t)$.}
\label{fig:function}
\end{figure}

We proceed now to show that for the purely electric case (I) of
spontaneously broken models ($v_{a}\neq 0$) $\rho $ is a strictly positive
function in all the domain of outer communications, $\langle \!\langle
\hbox{${\cal J}$\kern -.645em
{\raise.57ex\hbox{$\scriptscriptstyle (\ $}}}\rangle \!\rangle $, of a
static asymptotically flat black hole. In fact, let $\varepsilon >0$ be any
positive real number such that $0<\varepsilon \leq v_{1}$ (cf.~Fig.~\ref
{fig:poten}), where $v_{1}\neq 0$ is the least value minimizing the
potential function $U(\rho )$, then we shall show that $\rho \geq
\varepsilon >0$ in all of $\langle \!\langle
\hbox{${\cal J}$\kern -.645em
{\raise.57ex\hbox{$\scriptscriptstyle (\ $}}}\rangle \!\rangle $. This
result implies, by continuity of $\rho $, that $\rho \geq \varepsilon >0$
also at the horizon. In order to arrive at this conclusion, the equation of
motion (\ref{eq:scalar}) for $\rho $ will be used. Let $f_{\varepsilon }\in
C^{\infty }({\rm {I\!R}})$ be the real function defined by
\begin{equation}
f_{\varepsilon }(t)=\left\{
\begin{array}{cr}
-\exp (-1/(\varepsilon -t)^{2}), & t<\varepsilon , \\
0, & t\geq \varepsilon .
\end{array}
\right.  \label{eq:f}
\end{equation}
Such function satisfies the following conditions, cf.~Fig.~\ref{fig:function}%
,
\begin{equation}
f_{\varepsilon }(v_{a})=0,\qquad -1\leq f_{\varepsilon }(t)\leq 0,\qquad
f_{\varepsilon }^{\prime }(t)\geq 0,  \label{eq:prop}
\end{equation}
where $v_{a}$ is the value for which $U(\rho )$ achieves its $a$th minimum.

Multiplying Eq.\ (\ref{eq:scalar}) by $f_\varepsilon \circ \rho $ and
integrating by parts over ${\cal {V}}$, after applying the Gauss theorem,
one arrives at
\begin{equation}
\int_{\partial {\cal {V}}}f_\varepsilon (\rho )\nabla ^\mu \rho \,d\Sigma
_\mu =\int_{{\cal {V}}}\left( f_\varepsilon ^{\prime }(\rho )\nabla _\mu
\rho \nabla ^\mu \rho +\frac 12f_\varepsilon (\rho )U^{\prime }(\rho )+\frac{%
f_\varepsilon (\rho )}{4e\rho ^3}J_\mu J^\mu \right) dv.  \label{eq:ints}
\end{equation}
We would like to point out that the term $1/\rho ^3$ in the integrand above
is well behaved in the domain of outer communications $\langle \!\langle
\hbox{${\cal J}$\kern -.645em
{\raise.57ex\hbox{$\scriptscriptstyle (\ $}}}\rangle \!\rangle $. This rests
in the following: the integral identity (\ref{eq:ints}) is obtained from the
equation of motion (\ref{eq:scalar}). In order to this equation be satisfied
in the domain of outer communications $\langle \!\langle
\hbox{${\cal J}$\kern -.645em
{\raise.57ex\hbox{$\scriptscriptstyle (\ $}}}\rangle \!\rangle $, the
function $\rho $ must be $C^2\left( \langle \!\langle
\hbox{${\cal J}$\kern -.645em
{\raise.57ex\hbox{$\scriptscriptstyle (\ $}}}\rangle \!\rangle \right) $,
i.e., twice differentiable in this region. On the other hand, most of the
physically relevant potential are smooth functions, in fact, the mayor part
of them are polynomial. In this sense, the fulfillment of (\ref{eq:scalar})
implies, by the well--behaved nature of both its left hand side and the term
involving the derivative of the potential, that the remaining term in this
Eq., going as $1/\rho ^3$, is also well behaved in the domain of outer
communications $\langle \!\langle
\hbox{${\cal J}$\kern -.645em
{\raise.57ex\hbox{$\scriptscriptstyle (\ $}}}\rangle \!\rangle $.

In $\partial {\cal {V}}$ the boundary integrals over $\Sigma ^{\prime }$ and
$\Sigma $ cancel out again in the left hand side of (\ref{eq:ints}). The
boundary integral over $i^o\cap \overline{{\cal {V}}}$ vanishes, because $%
\rho $ takes asymptotically some of the values $v_a$ minimizing the
potential function $U(\rho )$, then by (\ref{eq:prop}) the integrand
vanishes there. The same happens to the integral over ${\cal {H}}^{+}\cap
\overline{{\cal {V}}}$; using the natural measure at the horizon the
integrand can be written as
\begin{equation}
f_\varepsilon (\rho )\nabla ^\mu \rho \,d\Sigma _\mu =f_\varepsilon (\rho
)\left( l_\mu \nabla ^\mu \rho +\left( l_\nu l^\nu \right) n_\mu \nabla ^\mu
\rho \right) d\sigma ,  \label{eq:integs}
\end{equation}
where the vanishing of $l_\mu \nabla ^\mu \rho $ and the bounded behavior of
$n_\mu \nabla ^\mu \rho $ at the horizon, together with the null character
of {\boldmath$l$} and the bounded behavior of $f_\varepsilon (\rho )$ (\ref
{eq:prop}) imply the vanishing of the whole integrand at the horizon. Since
there are no contributions at the left hand side of (\ref{eq:ints}), the
volume integral vanishes and we have for the purely electric case (I)
\begin{equation}
\int_{{\cal {V}}}\left( f_\varepsilon ^{\prime }(\rho )\gamma _{ij}\nabla
^i\rho \nabla ^j\rho +\frac 12f_\varepsilon (\rho )U^{\prime }(\rho )-V\frac{%
f_\varepsilon (\rho )}{4e\rho ^3}(J^t)^2\right) dv=0,  \label{eq:zeros1}
\end{equation}
where the coordinates from (\ref{eq:static}) has been used. From the
properties of $f_\varepsilon $, $U$, $V$, and {\boldmath$\gamma $} it
follows that each term in the integrand above is non--negative, so, (\ref
{eq:zeros1}) is fulfilled only if each of them vanishes identically in $%
{\cal {V}}$. In particular, $\left. f_\varepsilon (\rho )U^{\prime }(\rho
)\right| _{{\cal {V}}}=0$, this condition can be satisfied if $\left.
f_\varepsilon (\rho )\right| _{{\cal {V}}}=0$ which implies, from the
definition of $f_\varepsilon $ (\ref{eq:f}), that $\rho |_{{\cal {V}}}\geq
\varepsilon >0$. Conversely, let now suppose that $\left. f_\varepsilon
(\rho )\right| _p\neq 0$ for some $p\in {\cal {V}}$ this requires, from the
quoted condition, that $\left. U^{\prime }(\rho )\right| _p=0$ and, from the
definition of $f_\varepsilon $ (\ref{eq:f}), that $0\leq \rho
|_p<\varepsilon $, but the only extreme of $U(\rho )$ in this interval is at
$\rho =0$ (cf.~Fig.~\ref{fig:poten}), hence, $\left. f_\varepsilon (\rho
)\right| _p\neq 0\Rightarrow \left. \rho \right| _p=0$. The function $\rho $
can not vanishes in all of ${\cal {V}}$ because it asymptotically approaches
to one of the values $v_a$ for which $U(\rho )$ achieves its minima. Thus,
by the connectedness of ${\cal {V}}$ and the continuity of the function $%
\rho $, $\rho ({\cal {V}})$ is an interval in ${\rm {I\!R}}^{+}$ containing
the points $\left\{ 0,v_a\right\} $, which implies that the inverse image of
the open interval $\left] 0,\varepsilon \right[ \subset \rho ({\cal {V}})$
under the function $\rho $ is a nonempty open subset of ${\cal {V}}$; it is
clear that on this subset both $f_\varepsilon (\rho )$ and $U^{\prime }(\rho
)$ are nonvanishing functions (cf.~Figs.~\ref{fig:poten} and \ref
{fig:function}). Summarizing, the assumption $\left. f_\varepsilon (\rho
)\right| _p\neq 0$ for some $p\in {\cal {V}}$, implies the existence of a
nonempty open subset of ${\cal {V}}$ for which the condition $\left.
f_\varepsilon (\rho )U^{\prime }(\rho )\right| _{{\cal {V}}}=0$ is violated.
So, this contradiction implies the vanishing of $f_\varepsilon (\rho )$ in
all of ${\cal {V}}$, which requires, by the definition of $f_\varepsilon $ (%
\ref{eq:f}), that $\rho |_{{\cal {V}}}\geq \varepsilon >0$, result which can
be extended to all of $\langle \!\langle
\hbox{${\cal J}$\kern -.645em
{\raise.57ex\hbox{$\scriptscriptstyle (\ $}}}\rangle \!\rangle $. This
result finally implies, by the continuity of the function $\rho $ that $\rho
|_{{\cal {H}}^{+}}\geq \varepsilon >0$.

With the nonvanishing of $\rho $ at the horizon we have that (\ref{eq:int1v}%
) vanishes and (\ref{eq:int2b}) remains bounded in this region, which
implies, together with the null character of {\boldmath$l$} at the horizon,
the vanishing of the whole integrand (\ref{eq:integ}) over the horizon.

With no contribution from boundary integrals in (\ref{eq:int}), the volume
integral for the purely electric case (I) is written, using the coordinates
from (\ref{eq:static}), as
\begin{equation}
\int_{{\cal {V}}}-V\left( e^2\gamma _{ij}F^{ti}F^{tj}+\frac{4\pi }{\rho ^2}%
(J^t)^2\right) dv=0.  \label{eq:zeroE}
\end{equation}
The nonpositiveness of the above integrand, which is minus the sum of
squared terms, implies that the integral is vanishing only if $F^{ti}$ and $%
J^t$ vanish everywhere in ${\cal {V}}$, and hence in all of $\langle
\!\langle
\hbox{${\cal J}$\kern -.645em
{\raise.57ex\hbox{$\scriptscriptstyle (\ $}}}\rangle \!\rangle $.

Finally, we would like to explain why our proof on the nonvanishing of $\rho
$ fails in the purely magnetic case (II). This is due to the fact that the
last term in the volume integral (\ref{eq:zeros1}) must be replaced, in the
purely magnetic case (II), by the non--positive quantity $f_\varepsilon
(\rho )\gamma _{ij}J^iJ^j/4e\rho ^3$ (cf.~Fig.~\ref{fig:function}), since
the first two terms are again non--negative the integrand have no definite
sign and it is impossible to deduce the vanishing of it from the vanishing
of the integral. So, the nonvanishing of $\rho $ for the purely magnetic
case (II) must be justified using a different approach. We are looking for a
shortcut to solve this impasse, since we believe that the ``no--hair''
conjecture applies also to this case. For any successful justification of
the condition $\rho |_{{\cal {H}}^{+}}\neq 0$, the rest of the proof follows
in this way: the nonvanishing of $\rho $ implies again the vanishing of the
integrand (\ref{eq:integ}) over the horizon, having no contribution from
boundary integrals in (\ref{eq:int}), the volume integral for the purely
magnetic case (II) can be written, using the coordinates from (\ref
{eq:static}), as
\begin{equation}
\int_{{\cal {V}}}\left( \frac{e^2}2\gamma _{ij}\gamma _{kl}F^{ik}F^{jl}+%
\frac{4\pi }{\rho ^2}\gamma _{ij}J^iJ^j\right) dv=0,  \label{eq:zeroM}
\end{equation}
where again the non--negativeness of the integrand implies that the
vanishing of the integral is satisfied only if $F^{ik}$ and $J^i$ vanish
everywhere in ${\cal {V}}$, and hence in all of $\langle \!\langle
\hbox{${\cal J}$\kern -.645em
{\raise.57ex\hbox{$\scriptscriptstyle (\ $}}}\rangle \!\rangle $.

\section{``No--Hair'' Theorem for the Scalar Field in the Abelian Higgs Model
}

\label{sec:scalar}

It is reasonable to expect, from the ``no--hair'' conjecture, that the only
possible solutions for a scalar model in the domain of outer communications $%
\langle \!\langle
\hbox{${\cal J}$\kern -.645em
{\raise.57ex\hbox{$\scriptscriptstyle (\ $}}}\rangle \!\rangle $ of a
stationary asymptotically flat black hole become the vacuum expectation
values of the self--interaction. In the models considered in this paper this
implies the uniqueness of the scalar states $\Phi _a=v_a\exp i\theta $,
where $v_a\neq 0$ are the values minimizing the potential function $U(\rho )$%
. We now concentrate our attention in the Abelian Higgs model, for which $%
U(\rho )$ has a single minimum at $v$, and we shall show the truthfulness of
the last statement for the purely electric case (I), without any dependence
on the specific choice of the potential. The result is obtained by applying
the same procedure used above for the Eq.\ (\ref{eq:scalar}), with the
function $f_\varepsilon (\rho )$ replaced this time by the function $\tanh
(\rho -v)$, and taking into account that $J_\mu =0$, arriving now at
\begin{equation}
\int_{{\cal {V}}}({\rm sech}^2(\rho -v)\,\gamma _{ij}\nabla ^i\rho \nabla
^j\rho +\tanh (\rho -v)U^{\prime }(\rho ))dv=0,  \label{eq:zeros2}
\end{equation}
where the boundary integral vanishes by the same arguments yielding to the
vanishing of the boundary integral in (\ref{eq:ints}). Since $U(\rho )$ has
a single minimum at $v$, again the integrand at the left hand side of (\ref
{eq:zeros2}) is non--negative, so the integral vanishes only if $\rho =v$ in
all of ${\cal {V}}$, and hence in all of $\langle \!\langle
\hbox{${\cal J}$\kern -.645em
{\raise.57ex\hbox{$\scriptscriptstyle (\ $}}}\rangle \!\rangle $. We believe
that this result can be extended to more general Abelian models.

\section{Conclusions}

\label{sec:conclu}

The ``no--hair'' theorem for purely electric configurations of spontaneously
broken Abelian models has been extended to general static asymptotically
flat black holes. The theorem is gauge invariant, and is established for any
model with nonvanishing vacuum expectation values. It is shown that the
gauge field vanishes outside the black hole. This vanishing is physically
due to the effective behavior of the gauge field as a massive field by the
spontaneous symmetry breaking. For the particular case of the Abelian Higgs
model---Mexican--hat potential---it is additionally shown that the scalar
field is confined to the vacuum in all the black hole exterior, which
implies a zero contribution to the right hand side of the Einstein equations
(\ref{eq:Ein}), and that the only black hole admitted is the Schwarzschild
solution. We discuss the main conditions to establish the theorem for purely
magnetic configurations, but the problem remains still open; we believe that
the ``no--hair'' conjecture applies also to this case.

\acknowledgments
The author thanks Alberto Garc\'\i a for useful discussions and very
valuable hints, and Thomas Zannias for some considerations, in the early
stage of the work, about the correct measure that must be used in the
integrals at the horizon. This research was partially supported by the
CONACyT Grant 32138E and the Sistema Nacional de Investigadores (SNI). The
author also thanks all the encouragement and guide provided by his recently
late father: Erasmo Ay\'on Alayo.

\end{document}